\documentclass[article]{revtex4}

\usepackage{graphicx}
\usepackage{dcolumn}
\usepackage{bm}
\usepackage{color}
\usepackage{eso-pic}
\usepackage{amsmath }

\begin{document}

\large
\title{Phase preserving amplification near the quantum limit with a Josephson Ring Modulator}

\author{N. Bergeal$^{1,2}$, F. Schackert$^{1}$, M. Metcalfe$^{1,3}$, R. Vijay$^{1,4}$, V. E. Manucharyan$^1$, L. Frunzio$^1$, D. E. Prober$^1$, R. J. Schoelkopf$^1$, S. M. Girvin$^1$, M. H. Devoret$^1$}

\affiliation{$^1$Department of Physics and Applied Physics, Yale University, New Haven, CT, 06520-8284 USA.}
\affiliation{$^2$Laboratoire Photons Et Mati\`ere - UPR5-CNRS, ESPCI ParisTech, 10 Rue Vauquelin  75005 Paris, France.}
\affiliation{$^3$The Joint Quantum Institute, University of Maryland and the National Institute of Standards and Technology, College Park, MD 20742 USA.}
\affiliation{$^4$Department of Physics, University of California, Berkeley CA 94720-7300, USA.}

\date{\today}

\maketitle

Recent progress in solid state quantum information processing has stimulated the search for ultra-low-noise amplifiers and frequency converters in the microwave frequency range, 
which could attain the ultimate limit imposed by quantum mechanics.
In this article, we report the first realization of an intrinsically phase-preserving,  non-degenerate superconducting parametric amplifier, a so far missing component. It is based on the Josephson ring modulator, which consists of four junctions in a Wheatstone bridge configuration. The device symmetry greatly enhances the purity of the amplification process and simplifies both its operation and analysis. The measured characteristics of the amplifier in terms of gain and bandwidth are in good agreement with analytical predictions. Using a newly developed noise source, we also show that our device operates within a factor of three of the quantum limit. This development opens new applications in the area of quantum analog signal processing.\\

In this article, we focus on parametric amplifiers which are powered by an ac source with frequency $f_p$ also known as the ``pump". Such amplifiers operate with a minimal number of degrees of freedom and are the natural candidates for ultra low noise operation \cite{louisell, gordon}.
A single spatial and temporal mode of the electromagnetic  field with carrier frequency $f$ can be decomposed into its in-phase $A_{\parallel}\cos2\pi ft$ and out-of-phase  $A_{\perp}\sin2\pi ft$ quadratures. Linear amplifiers can be classified into two categories from the way they treat these two quadratures. On one hand, a phase-preserving amplifier treats both quadrature components with the same gain ($\sqrt{G}$) as an ordinary op-amp would do ($A_{\parallel}\rightarrow\sqrt{G}A_{\parallel}$ ; $A_{\perp}\rightarrow\sqrt{G}A_{\perp}$). On the other hand, in a phase-sensitive amplifier, the gains of the quadratures are inverse of each other ($A_{\parallel}\rightarrow\sqrt{G}X_{\parallel}$; $A_{\perp}\rightarrow\frac{1}{\sqrt{G}}A_{\perp}$). Regarding their noise properties, there is a fundamental difference between these two types of amplifiers. The minimum noise energy added by a phase-preserving amplifier to the input signal  amounts to half a photon at the signal 
frequency $\frac{1}{2}hf $, where $h$ is Planck's constant \cite{haus,caves,clerk}. On the other hand, a phase-sensitive amplifier is submitted to  only a lower limit on the product of the noise added to the two quadratures and can thus squeeze the quantum noise on one quadrature at the expense of extra noise in the other \cite{caves,castellanos}. Although such amplifiers can look rather appealing because of their ability to operate potentially below the quantum limit, they are only useful in cases where a reference phase is attached to the signal, like in homodyne detection.  In  the majority of cases, where the information carried by the signal is contained in both quadratures, or equivalently in both amplitude and phase, a phase-preserving amplifier is preferable. However, so far, little attention has been devoted to non-degenerate, intrinsically phase-preserving, parametric amplifiers operating near the quantum limit in the microwave frequency range. They involve two  distinct internal resonant  modes of the circuit, conventionally called the ``signal''  with center  frequency $f_{s}$ and the ``idler'', whose frequency $f_{i}$ differs from $f_s$ by at least the sum of the bandwidth of the two resonances. This is in contrast with phase-sensitive degenerate parametric amplifiers which operate with one internal resonant mode only \cite{lehnert, yamamoto, yurke88,yurke89,movshovich}.
The challenge of building phase-preserving amplifiers arises from the difficulty of controlling the two modes and coupling them together with the pump.\\
  \begin{figure}[htbp]
\centering
\includegraphics[width=12cm]{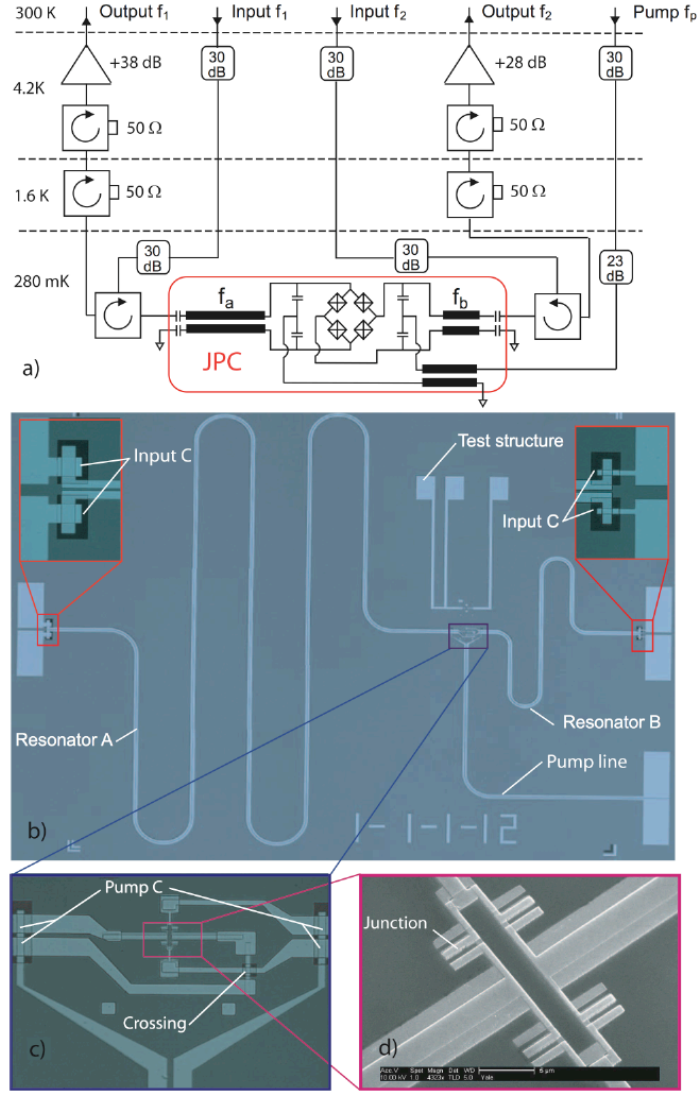}
\caption{The Josephson Parametric Converter and its microwave measurement set-up. a) JPC sample (red  solid line box) whose signal ports are connected to two input lines and two output lines using cryogenic circulators, while its  pump  port is  fed by a fifth line. b) Optical  picture of the JPC. It consists of two CPS  Al resonators of length 17 mm and 3.6 mm  coupled to the Josephson ring modulator on one side and to contact pads via input capacitances on the other sides. A third CPS line carries the pump signal. Input capacitances are built with two Al layers separated by
a SiO$_x$ dielectric layer. c) Closeup of the connection between the Josephson ring modulator, the resonators and the pump line. The pump is weakly coupled to the resonators through the SiO$_x$ dielectric layer which provides a small capacitance of order 20 fF. The crossing point of the two resonators is isolated  by the same SiO$_x$ layer. d) SEM pictures of the Josephson ring modulator showing its four Al/AlO$_x$/Al junctions. Each junction is surrounded by the two shadow electrodes produced by the Dolan bridge double angle evaporation  technique. The loop of the ring is 3 $\mu$m $\times$ 17 $\mu$m and the junction area is 5$\mu $m $\times$1 $\mu$m. }
\end{figure} 

\textbf{Description of the Josephson Parametric Converter}\\

The circuit nicknamed Josephson Parametric Converter (JPC) has been theoretically described in reference \cite{bergeal}. A schematic description can be found in the red solid line box of Fig. 1a.  Its operation is based on a novel non-linear device, the Josephson ring modulator, consisting
 of four nominally identical Josephson junctions forming a superconducting loop \cite{bergeal}.
It has the minimal number of electromagnetic spatial modes required to perform 3-wave mixing:  two differential ones, called $X$ and $Y$, and a common one, called $Z$. In the JPC, two superconducting resonators, A and B, couple to the differential modes $X$ and $Y$ of the ring, respectively. In the following,  $f_{a(b)}$  and $Q_{a(b)}$ denote the resonant frequencies and  quality factors of the resonators. An additional transmission line carries the pump signal at frequency $f_p$, and is weakly coupled to the common mode $Z$ of the ring through capacitances.  From the point of view of the signals, the device has two ports:  port 1,  driven at  frequency $f_{1}$, feeds resonator A and port 2, driven at  frequency $f_{2}$, feeds resonator B. In contrast with previous Josephson parametric amplifier, we have here a complete separation of the signal and idler modes both spatially and temporally. The JPC can be operated as a phase-preserving amplifier when $f_p=f_{1}+f_{2}$ or as a unity photon gain frequency converter when $f_p=|f_{2}-f_{1}|$.  In this paper, we will
 focus only on the amplification mode, in which case the JPC is described by the input-output relation  \cite{bergeal}

\begin{eqnarray}
\hat{a}^{_{\mathrm{out}}}_{1}=r_1\hat{a}^{_{\mathrm{in}}}_{1}+s_1\hat{a}^{\dagger_{\mathrm{in}}}_{2}\\
\hat{a}^{_{\mathrm{out}}}_{2}=r_2\hat{a}^{_{\mathrm{in}}}_{2}+s_2\hat{a}^{\dagger_{\mathrm{in}}}_{1}
\label{matrix1}
\end{eqnarray}

The $\hat{a}_{1(2)}$ and $\hat{a}_{1(2)}^\dagger$ are annihilation and creation operators at port 1 (frequency $f_1$) and port 2 (frequency $f_2$), expressed in units of the square root of photon number per unit time.  Relations (1)  and (2)  thus determine the properties of the amplifier including its noise in the quantum limit.\\

The  coefficients $r_{1(2)}$ and $s_{1(2)}$ satisfy the symplectic relation  $|r_{1(2)}|^2-|s_{1(2)}|^2=1$. They are given by 
\begin{eqnarray}
r_{1(2)}=-\frac{(\vartheta _{2(1)}+i)(\vartheta _{1(2)}+i)-|\rho |^{2}}{(\vartheta _{2(1)}+i)(\vartheta _{1(2)}-i)-|\rho |^{2}} \quad \text{and} \quad s_{1(2)}=\frac{-2i\rho }{(\vartheta _{2(1)}+i)(\vartheta
_{1(2)}-i)-|\rho |^{2}}
\label{ex_gain}
\end{eqnarray}

where  $\vartheta_{1}=Q_{a}\frac{f_{a}^2-f_{1}^2}{f_{1}f_{a}}$, and $\vartheta_{2}=Q_{b}\frac{f_{b}^2-f_{2}^2}{f_{2}f_{b}}$.\\

  \begin{figure}[htbp]
\centering
\includegraphics[width=10cm]{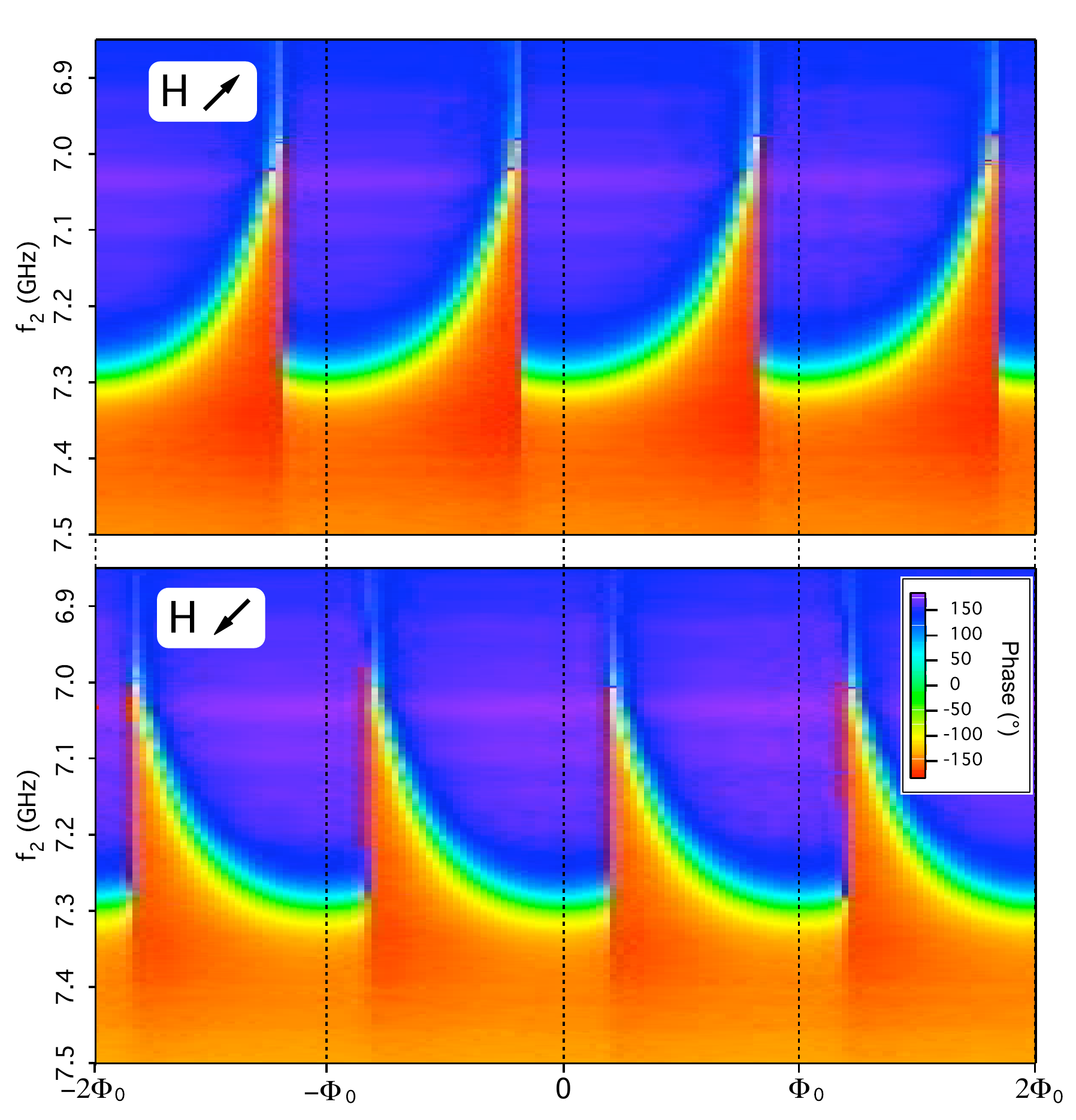}
\caption{  Modulation of  resonance frequency with magnetic field. Phase of the reflected signal at port 2 (color), as a function of the drive frequency 
and magnetic flux through the loop, for two directions of the magnetic field sweep.}
\end{figure}  
 
Here, we have introduced the dimensionless pump current $|\rho|$ 
 \begin{eqnarray}
|\rho|=\frac{1}{4}\sqrt{Q_{a}Q_{b}p_{a}p_{b}}\frac{I_{p}}{I'_{0}}
\label{rho}
\end{eqnarray}
where $I_p$ is the pump current, $I'_{0}$ is the Z mode critical current and  $p_{a(b)}=\frac{L_J}{L_J+L_{a(b)}}$ are the participation ratios of the inductance of the Josephson ring modulator $L_J$ to the resonator inductances  $L_{a(b)}$ \cite{bergeal}. At the resonant tuning $f_{1}=f_{a}$ and $f_{2}=f_{b}$, the expressions of the coefficients reduce to $r_1=r_2=\sqrt{G}=\frac{1+|\rho|^2}{{1-|\rho|^2}}$ and $s_1=s_2=-i\sqrt{G-1}=\frac{-2i\rho}{{1-|\rho|^2}}$. The symbol $G$ refers to
the power gain at the band center.

The diagonal term $r_{1(2)}$ can be seen as a photon ``cis-gain'', characteristic of the 1-port reflection amplifier operation : the incoming wave at either port is 
reflected with a power gain $|r_{1(2)}|^2$ and its phase is preserved.
The non-diagonal term $s_{1(2)}$ can be seen as a photon ``trans-gain'' between the two ports, and describes the behaviour of a frequency converter ($f_{1}\rightarrow f_{2}$ or $f_{2}\rightarrow f_{1}$) with power gain $|s_{1(2)}|^2=|r_{1(2)}|^2-1$.  The possibility of performing the important function of up- and down-conversion is another feature that distinguishes the non-degenerate parametric amplifier from the degenerate one. It arises from the presence in the circuit of two different spatial modes with different frequencies, whose  sum is equal to the pump frequency. The  ``cis-gain'' and ``trans-gain'' can be varied simply by changing the pump power through the $|\rho| $ parameter, $|r_{1(2)}|\simeq|s_{1(2)}|\underset{|\rho| \to 1^-}{\longrightarrow}\infty$.\\ 
In the large gain limit, the expressions of $|r_{1,2}|$ and $|s_{1,2}|$  reduce to a Lorentzian form
\begin{eqnarray}
|r_{1,(2)}|\simeq |s_{1(2)}|\simeq\frac{\sqrt{G}}{\sqrt{1+G\big(\frac{Q_a}{f_a}+\frac{Q_b}{f_b}\big)^2\big(f_{1,2}-f_{a,b}\big)^2}}
\end{eqnarray}
The -3dB bandwidth of the amplifier is thus
\begin{eqnarray}
B=\frac{2}{\sqrt{G}}\Big(\frac{Q_a}{f_a}+\frac{Q_b}{f_b}\Big)^{-1}
\label{ex_b}
\end{eqnarray}

  \begin{figure}[htbp]
\centering
\includegraphics[width=12cm]{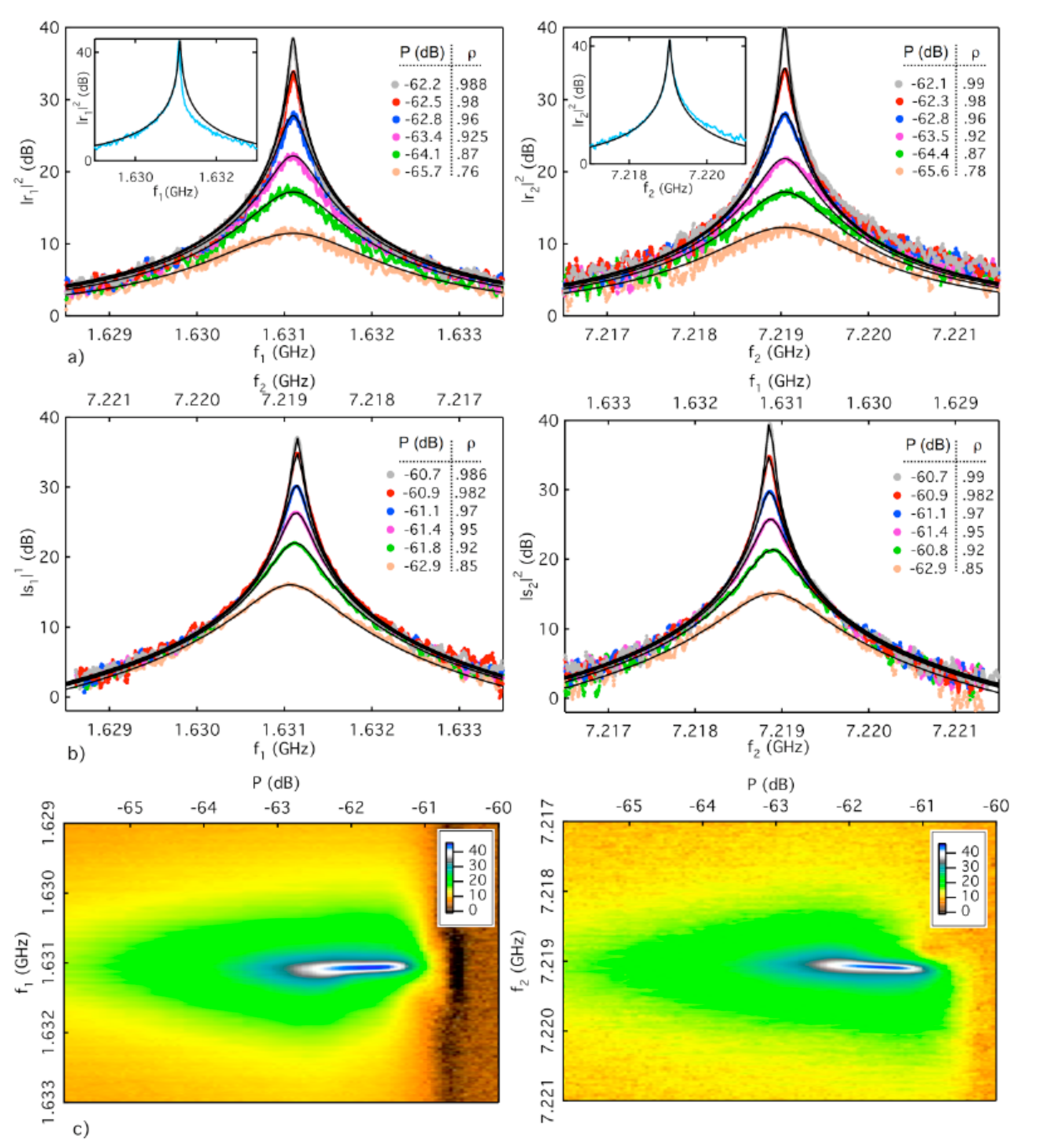}
\caption{ Gain of the JPC.  a) Power ``cis-gain" of the JPC as a function of the drive frequency for different values
 of  the pump power measured at  port 1  (left) and port 2 (right).  The solid lines correspond to the theoretical expressions of $|r_1|^2$ and $|r_2|^2$ (\ref{ex_gain}) 
    obtained for the different values of the fit parameter $|\rho|$. Inset: curves obtained at larger gain. The fits correspond to  $|\rho|=0.994$ (left) and
  $|\rho|=0.992$ (right). b) Photon ``trans-gain"  of the JPC as a function of the drive frequency (bottom axis) and the converted frequency (top axis) for different values  of the pump power measured between  port 1  and  port  2 (left) and between  port 2 and port 1 (right). The solid lines correspond 
 to the theoretical expressions $|s_1|^2$  and $|s_2|^2$ (\ref{ex_gain}) obtained for the different values of
  the parameter $|\rho|$.  c) Photon ``cis-gain"  of the JPC plotted in color scale as a function of the drive frequencies and the pump power measured at port 1(left) and port 2 (right). Note that the data shown in the different panels a, b, c correspond to different runs.}
\end{figure}

 \textbf{Implementation and characterisation of the JPC}\\
 
 Fig. 1b-d  show  pictures of our JPC sample made of Aluminium on a Si wafer, using a three-step fabrication process combining optical and e-beam lithography. The details of fabrication are given in the Methods section. 
 The two resonators  A and B are 17 mm  and 3.6 mm long. The four junctions of the ring have nominally identical critical currents with values in the range 3-6$\mu $A. Tests have shown that these values  vary from one junction of the ring to another by less than 5 $\%$. The center frequencies of the resonators are slightly renormalized by the Josephson inductance  $L_J$ of the ring and are, in absence of flux through  the ring,  $f_a$=1.6 GHz and $f_b$=7.3 GHz. The choice of these frequencies results from a compromise between separation of signal and idler for frequency conversion, and complexity of microwave
 engineering. Our resonators are both in the overcoupled regime and the quality factors are therefore determined by the overlap coupling capacitances connecting the resonators to the external circuit. Here, the values $Q_{a}$=450 and $Q_{b}$=120 result from favouring capacitor reliability and have not
 been optimized for maximum amplifier performances.  
 After  fabrication, the JPC chip is mounted on a microwave circuit board and housed in a  copper 
 sample box equipped with a superconducting coil.  The sample box is anchored to the 280 mK cold stage of a $^3$He refrigerator and placed in a magnetic shield.  Although similar embedding of  $\mu$A critical current junctions inside microwave resonators have recently been employed for bifurcation readout of qubits \cite{mike,boulant} and RF magnetometry \cite{vijay}, the ring modulator device investigated here works in the weakly non-linear regime,  with the pump amplitude well below the bifurcation threshold.  In other words, the currents in the ring modulator remain small enough that only second order correction to the Josephson inductance need to be considered.\\
 A schematic description  of our measurement set-up is shown in Fig. 1a. Two attenuated input lines carry the signal from the microwave generators 
 to the JPC. Circulators separate  the output signal reflected by the JPC from the input signal at the coldest stage. The output signals are first amplified by HEMT cryogenic amplifiers at the 4.2K stage. Two isolators are placed at  the 4.2K and 
 1.6K stages to minimize the backaction of the amplifier on the sample. In this arrangement, the incoming noise  at the input ports of the JPC arises from the 50 $\Omega $ resistance presented by the last
 attenuator in the input lines. At room temperature, the signal is further amplified  by about 60 dB before being measured. 
  The frequency bands of  the output lines are limited by the isolators, circulators and amplifiers: 1.2-1.8 GHz for the low frequency ($f_1$) line  and 4-8 GHz for the high frequency ($f_2$) line.\\

  \begin{figure}[htbp]
\centering
\includegraphics[width=10cm]{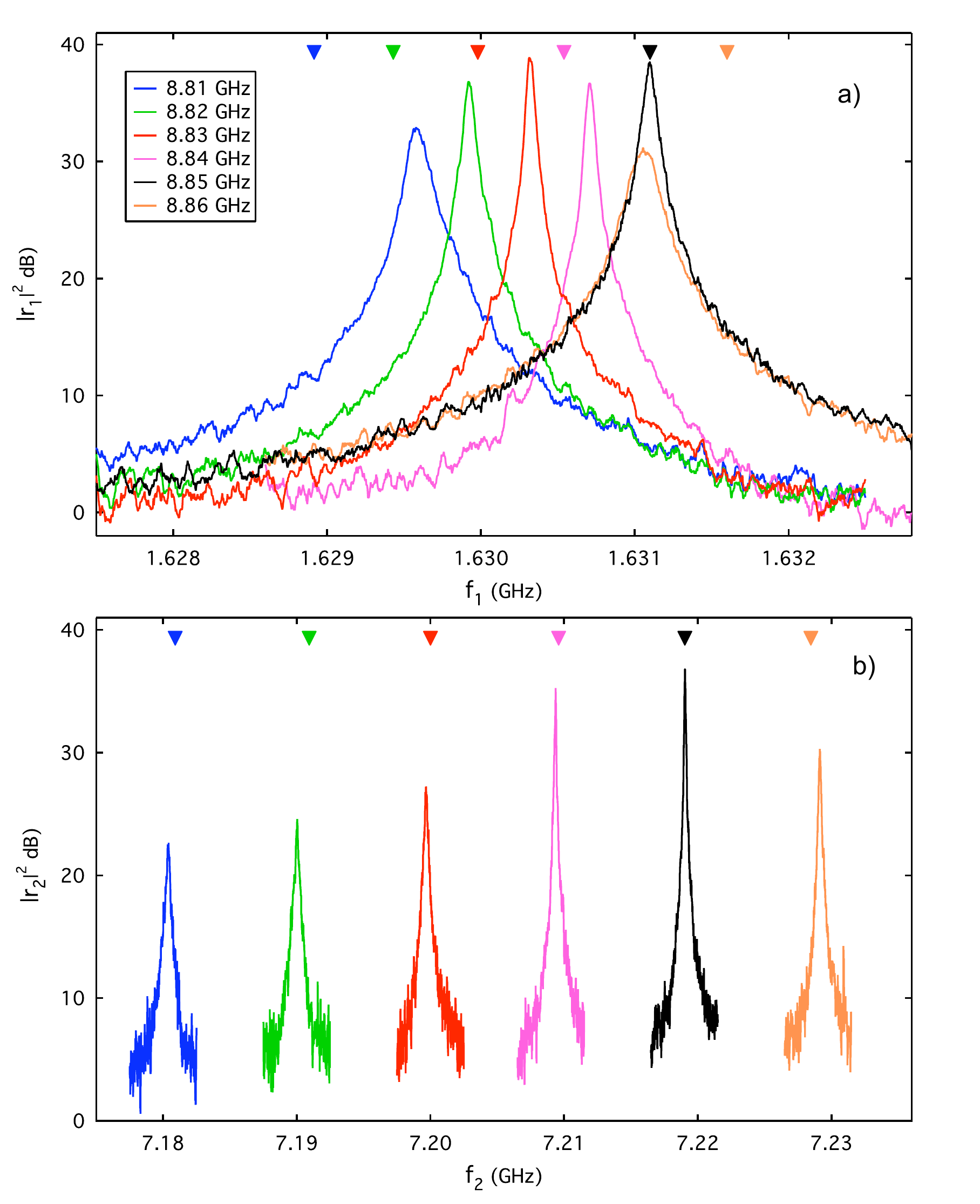}
\caption{ ``Cis-gain" $|r_{1(2)}|^2$ at port 1 (top) and port 2 (bottom) as a function of drive frequencies for different values of the pump frequency $f_p$. The pump amplitude has been adjusted at each frequency for optimal gain. The triangular symbols indicate the theoretical location of the center frequency. }
\end{figure} 
     
We characterized our resonators with a vector network analyzer.  In the reflection measurements, there was no amplitude response and the resonance frequencies, showed up only in the phase response which displayed a complete $2\pi$ phase shift, confirming the completely dispersive nature of the device. Then, we followed the phase response as a function of the dc current applied to the coil which is proportional to magnetic flux $\Phi$ through the ring.
 As expected, we observe $\Phi_{0}$-periodic modulations of the resonance frequency (Fig. 2) which are hysteretic with the magnetic field ($\Phi_{0}$ denotes the superconducting flux quantum). According to Fig. 1b of reference \cite{bergeal}, the two lower energy states cross each other for a  flux bias $\frac{\Phi_{0}}{2}$. 
However, as the two  states
are separated by an energy barrier, the system initially in the ground state for $\Phi=0$ becomes metastable when the flux is increased beyond 
$\frac{\Phi_{0}}{2}$.  As the  barrier is maximum at the degeneracy point and decreases to zero at $\Phi_{0}$, the system remains in the 
metastable state until its environment (noise, measurement signal) provides a sufficient amount of energy to activate the transition. The operation of the JPC  
is optimal when biased in the vicinity of $\frac{\Phi_{0}}{2}$ \cite{bergeal}. In the following, we will focus
on measurements performed at this bias point where the resonance frequencies where found to be $f_{a}$=1.631 GHz and $f_{b}$=7.219 GHz.\\
    \begin{figure}[htbp]
\centering
\includegraphics[width=10cm]{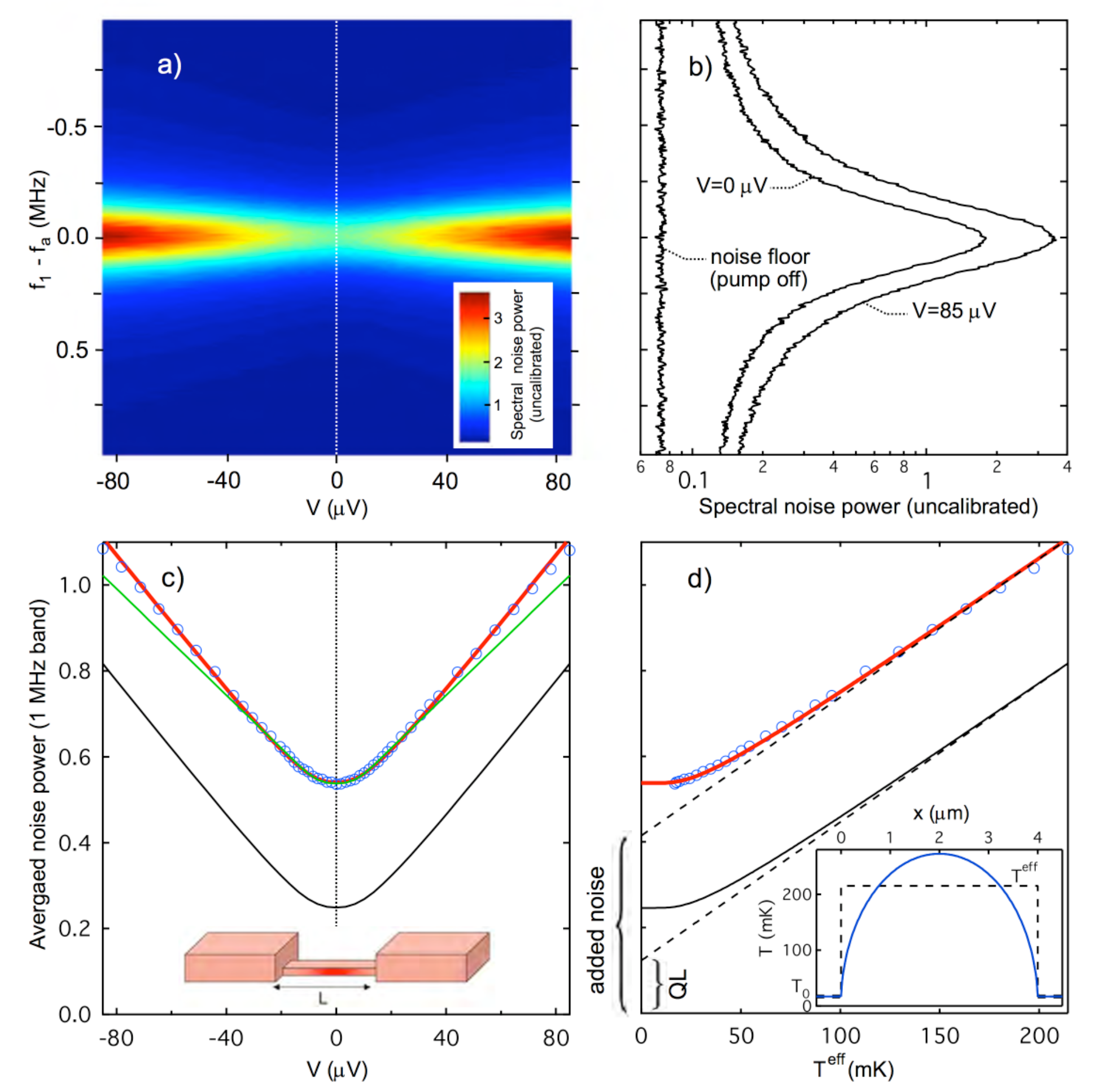}
\caption{Noise measurement of the JPC for a 30 dB gain. a) Spectral noise spectrum at the output of port 1 (color) as a function of frequency and voltage across the resistor. b) Cuts of the noise power spectrum corresponding to V=0 $\mu$V and V=85 $\mu$V. The noise floor obtained with pump off is given as reference. c) Blue open dots: noise power at the ouput of port 1, same units as in a), averaged in a 1MHz band around $f_a$, as function of the voltage across the resistor.  Red line: theoretical expression (\ref{noise}) with fitted added noise; green line: same as red line but assuming quantum shot noise for the resistor; black line: theoretical expression (\ref{noise}) and an ideal quantum limited amplifier.  Inset: schematic description of the resistor embedded in its thick and wide thermal reservoirs. d) Same as c) but with the voltage axis converted into the effective temperature of the noise source $T^{\mathrm{eff}}$.
The dashed lines indicate asymptote of the high temperature variation. Also shown in this panel is the total added noise power of the system and the ideal case
of the quantum limit (QL). Inset shows the temperature variation profile inside the resistor (see Methods). }
\end{figure} 
      
  \textbf{Operation of the JPC}\\
   
 The device was operated in the amplifier mode by feeding the pump line with a continuous microwave signal at 
 a fixed frequency $f_p=f_{a}+f_{b}$ = 8.85 GHz.  In this mode, the JPC is described by relations (1) and (2).
   Fig. 3a shows the gain curves $|r_{1(2)}|^2$ measured at both ports for different values of the pump power.  They are symmetrically 
 centered on the resonance  frequencies $f_{a}$ and $f_{b}$ and, importantly, the symmetry of the ring modulator is such that they do not shift noticeably with the pump power, unlike in other Josephson parametric amplifiers \cite{lehnert,yamamoto,yurke89}. In addition to the experimental data, we also plot
  the theoretical expression (\ref{ex_gain}) for  $|r_{1(2)}|^2$ as a function of drive frequencies $f_{1(2)}$, using the 
 values of  $f_{a}$, $f_{b}$, $Q_{a}$ and $Q_{b}$ measured previously. Data and theory are in very good
  agreement up to 40 dB. This is quite  remarkable since only one adjustable parameter $|\rho|$ is sufficient to capture the whole shape of the gain curves. In particular  the relationship (\ref{ex_b}) between gain and bandwidth is well satisfied.   A similar agreement with theory is also obtained for the phase of the complex gain $r_{1(2)}$ as shown in  supplementary figure 1.
 As far as the relationship between $|\rho|$ and $I_p$ is concerned, we found experimentally that, for a wide range of gain, $|\rho|\propto \mathrm{log}I_p$. This dependence differs from the linear relation (\ref{rho}) obtained in the framework of reference  \cite{bergeal} assuming a stiff pump, an ideal situation  difficult to realize in practice since the coupling of the pump mode changes as the gain increases.
 When the gain is further increased to a maximum of 44 dB, the curves deviates from the model as reported in insets of Fig. 3a, showing that  saturation effects are  becoming important, as  discussed below. These effects occur within a fraction of a percent  from the theoretical onset of self-oscillations of
 the system corresponding to  $|\rho|=1$. Moreover, as shown in Fig. 3c, after reaching its maximum value the gain starts to decrease when further increasing the pump power and finally collapses when the critical current of the junctions is reached. We attribute the failure to observe parametric self-oscillations to the saturation of the device by the signal resulting from the amplification of the noise, whose peak amplitude reaches the critical current.

 A practical amplifier must have a minimum gain to beat the noise of the following amplifier in the measurement chain. Assuming the best  ``state-of-the-art'' commercial device as a following amplifier with a noise temperature of a few K, we are led to a requirement of at least 20 dB of power gain, satisfied
with a comfortable margin by the present JPC. This is in contrast with recent results obtained with other parametric amplifiers \cite{lehnert, yamamoto} or dc SQUID microwave amplifiers \cite{spietz,kinion}. We believe that the differences in gain performances between our JPC and other Josephson amplifiers originate in the purity of the non linear $XYZ$ coupling of signal, idler and pump offered by the ring modulator. Indeed, the observed maximum gain is compatible with the order-of-magnitude theoretical prediction given by

\begin{eqnarray}
G<\frac{I_0}{2\pi e(p_af_a+p_bf_b)}
\label{limit}
\end{eqnarray}
 where the participation ratio $p_a$ and $p_b$ are  0.02 and 0.09 respectively and $I_0$ is  3-6 $\mu$A  \cite{bergeal}.\\
 The same consideration based on the maximum output power that the Josephson ring modulator can deliver also explains the dynamic range of our amplifier (supplementary figure 2).
  In addition to the 1-port amplifier operation corresponding to the diagonal terms $r_{1(2)}$, the JPC also performs 2-ports frequency 
   conversion with gain corresponding to the non-diagonal terms $s_{1(2)}$. Fig. 3b shows the typical gain curves measured between the two ports both in the cases of 
    up conversion ($f_{1}\rightarrow f_{2}$) and down conversion ($f_{2}\rightarrow f_{1}$). Like with $|r_{1(2)}|$ we obtain an excellent agreement with the theoretical expression of $|s_{1(2)}|$.

    In Fig. 4, we show that  we can adjust the center frequency of the signal bandwidth by detuning the pump frequency. Thus, the tuning bandwidth is found to be given 
    by the intrinsic resonator bandwidth $\frac{f_{a(b)}}{Q_{a(b)}}$ as expression (\ref{ex_gain}) predicts. For the high frequency port of our device, there is a ratio of 95 between the tuning and signal bandwidth if we limit the gain to 20 dB. For the low frequency port, this ratio is much smaller (5.7) since $\frac{f_{a}}{Q_{a}}\ll\frac{f_{b}}{Q_{b}}$. Finally, let us note that the relatively small signal bandwidth of the present device is due to the small intrinsic bandwidth of the low frequency resonator and could be
    greatly enhanced by increasing its coupling capacitance and the working frequency. The optimum bandwidth configuration occurs when the signal and idler bandwidth are equal. \\

   \textbf{Noise measurement}\\

 To measure the noise of the JPC we have developed a  source of noise based on the hot-electron shot noise regime of a $R$=50$\pm$1 $\Omega$ mesoscopic copper
 resistor of length $L=4\mu$m embedded between two cold reservoirs (inset Fig. 5a). It is inspired by a previous work in which this system was used to calibrate
 an infrared photon detector \cite{teufel}.  A detailed description 
 of the device  is given in the Methods section. This noise source has the following advantages : (i)  no macroscopic heating of the sample and thus no perturbation
 of the JPC operating point  (ii) self-calibration  (iii) response time in the microsecond range  \cite{teufel} (iv) control by a dc current $I_{dc}$. 
The Johnson noise spectral power produced by the resistor is given by
 \begin{eqnarray}
S^{ns}(f,V)=\frac{hf}{2}\coth\bigg(\frac{hf}{2k_{\mathrm{B}}T^{\mathrm{eff}}(V)}\bigg)
\label{noise}
\end{eqnarray}
where $T^{\mathrm{eff}}$ is the effective electronic temperature of the resistor, a function of the voltage $V=RI_{dc}$ across the resistor, and 
$k_{\mathrm{B}}$ the Boltzmann constant.
 For $k_\mathrm{B}T^{\mathrm{eff}}\gg hf$ this expression reduces to the well known classical Johnson spectral noise power $k_\mathrm{B}T^{\mathrm{eff}}$.\\

In this experiment, both the noise source and the JPC were  anchored to the mixing chamber of a dilution refrigerator with base temperature $T_{0}=17$ mK and connected
to port 1 via a circulator (supplementary figure 3). Fig. 5a shows the noise in color as a function of frequency and voltage $V$, with the gain of the JPC  set to 30 dB. A cut of data taken
at V=0 $\mu$V and V=85 $\mu$V is shown is Fig. 5b in dB scale.  Fig. 5c  shows the noise averaged  on a 1MHz band around the resonance frequency.
In the same figure, we compare the data with the theoretical expression 
 \begin{eqnarray}
S^{th}=G_{t}\big( S^{ns}(V,f_1)+S^{\mathrm{add}}\big)
\end{eqnarray}
where $G_{t}$ is the total gain of the measurement chain and  $S^{\mathrm{add}}$ is the noise added by the amplifier. In the case
of a perfect quantum limited (QL) amplifier, $S^{\mathrm{add}}=S^{QL}=\frac{hf_{1}}{2}$, which corresponds to the noise brought in by the idler when the load at port 2 is at a temperature $T_0$ such that $hf_2\gg k_\mathrm{B}T_0$. Note that the vertical axis in Fig 5c  is the total noise of the system. Yet, it entirely measures the output noise of the JPC since the noise added by the following stages is 17 dB lower, as shown in Fig 5b.
There is a good agreement between theory and experimental data if $S^{\mathrm{add}}=3.37\times S^{\mathrm{QL}}$ (red curve). In the same figure, for comparison we  plot the theoretical result corresponding to the ideal case of both perfect nanowire and amplifier (black curve). The quality of the fit confirms that the resistor is well in the hot electron regime. For instance, the expression for  quantum shot noise (green curve in Fig. 5c, see Methods) cannot been reconciled with the experimental data. In Fig. 5d
  the voltage axis has been converted into a temperature scale using the expression for $T^{\mathrm{eff}}$ (see Methods). In this figure, we can read directly
  the noise added by the amplifier as the vertical intercept of the dashed line corresponding to the asymptotic variation of noise with temperature. Converting the added noise $S^{\mathrm{add}}$ into the system noise temperature, we found $T_{N}=125 $ mK which makes the JPC 20-40 times better than the best commercial HEMT amplifier. We attribute the difference between the system noise temperature and the ideal quantum limit $T_N=37$ mK to incomplete isolation of the JPC from spurious noise coming through the post-amplifiers and possibly to imperfections in the line connecting the nanowire to the JPC, which incorporates a circulator and a bias tee. \\

In conclusion, we have implemented and demonstrated the operation of a new phase-preserving amplifier, the Josephson Parametric Converter, whose
main element is the nonlinear, nondissipative inductance of a Josephson ring modulator. Our results concerning the gain, bandwidth and dynamic range
can be  understood quantitatively using a simple analytical model involving a minimum number of parameters. We have achieved a power gain amplification 
of at least 40 dB,  which opens the possibility of practical applications like single photon detection based on microwave readout \cite{dan} or any applications involving the measurement of very low microwave power in a limited bandwidth. Our device is also useful as a  frequency up-or-down converter. Using a newly developed self-calibrating noise source,
we obtained an upper bound on the noise added by the JPC which is three times the quantum limit.
Extrapolating the present results to an optimal choice of parameters, we should be able to build 
a quantum limited amplifier with 20 dB gain in a 10 MHz bandwidth, allowing dispersive single shot readout of solid state qubits with an irradiation of only a few photons \cite{wallraff,majer, lupascu,sillamaa,vlad}. Our results also open the way for production of entangled microwave signal pairs through two-mode squeezing operation.\\

Correspondence and requests for material should be address to M. H. D. and N. B.\\

 This work was supported by NSA through ARO Grant No. W911NF-05-01-0365, the Keck
foundation, and the NSF through Grant No. DMR-032-5580. M. H. D. acknowledges partial support from College de France.

 \newpage
 \textbf{Methods}\\
 \textbf{Fabrication of the device}\\
 
Each quarter wave resonator is built from a Coplanar Strip Line (CPS)  structure consisting of two 15 $\mu$m wide parallel 
 lines separated by a 4 $\mu$m gap. In the first step of fabrication, a 110 nm thick Al
layer is evaporated onto a high resistivity Si wafer through an optical bilayer resist. 
  A slope edge is obtained by tilting the sample holder with 
 an angle of 4$^\circ$ and rotating the stage at the speed of 10$^\circ$/s during the evaporation. This process defines the CPS resonators and the bottom layer of the input capacitances.
 Then a 270 nm thick SiO$_x$ layer is deposited through a second
 optical  resist to form in a single step, the dielectric layer of the input capacitances, the pump coupling capacitances and the isolation at the crossing of the two resonators line. Finally 
 a second layer of Al (240 nm) is evaporated to  form  the pump line, the top layer of the input capacitances connected to contact pads, and the bridge on the top of the isolation. 
  After the optical processes, the wafer is diced into 50 chips of size 5$\times$4 mm$^2$. Next, the Josephson  ring modulator is
integrated  to the device using electron beam lithography, MMA/ PMMA resist bilayer and the Dolan bridge double angle evaporation technique. A gentle hollow cathode ion gun milling was employed  between deposition of metallic layers. 
 The nominal size of each junction is 5$\times$1 $\mu$m$^2$  yielding a resistance of tunnel resistance of  40/80$\Omega$ depending on oxidation conditions.\\

 \textbf{ Noise measurement with a mesoscopic resistor }\\
 
  A metallic resistor exhibits several regimes of shot noise depending on the relation between its length $L$  and the characteristic length scales involved in the motion of electrons \cite{steinbach}.
In the regime where $L$ is longer than the inelastic electron scattering length $L_{e-e}$, but shorter than the electron-phonon interaction length $L_{e-ph}$, the electrons  travelling through the resistor redistribute the excess kinetic energy gained from the voltage via electron-electron interactions only \cite{steinbach}. 
The cooling of the electrons system thus occurs only by diffusion to the cold reservoirs and is controlled by the Wiedemann-Franz law. 
As a result, an equilibrium electronic temperature profile establishes itself over the length of the microbridge (inset Fig. 5b) \cite{nagaev}
\begin{eqnarray}
T(x)=\sqrt{1+\frac{x}{L}\big(1-\frac{x}{L}\big)\frac{3e^2V^2}{\pi^2k^2_{B}T_{0}^2}}
\end{eqnarray}

An effective temperature for the nanowire can be obtained by
 integrating over its length $T^{\mathrm{eff}}=\frac{1}{L}\int^L_{0}T(x)=\frac{T_{0}}{2}\Big\lbrack1+\Big(v+\frac{1}{v}\big)\arctan(v)\Big\rbrack$ 
where $v=\frac{\sqrt{3}}{2\pi}\frac{eV}{k_\mathrm{B}T_{0}}$.  We have verified that when computing the effect of shot noise, using this effective temperature and the
formula for the Johnson noise gave, at the percent precision level, the same result as when the full space dependent  spectral density of noise is taken into account.\\ 
 Our  50 $\Omega$ Cu nanowire resistor is 4 $\mu$m long, $80$ nm wide, and $20$ nm thick. Using a double angle evaporation technique, the resistor is 
 embedded into a  500 nm thick 50 $\Omega$ CPW transmission line of  400 $\mu$m width providing good thermal reservoirs.  The noise source chip is bonded
 with gold wire on a sample holder itself thermally  anchored to the mixing chamber of the dilution fridge with base temperature $T_{0}$=17 mK.
  According to previous studies \cite{pothier}, the 4 $\mu$m length  ensures that the resistor is in the
hot electron regime $L_{e-e}<L<L_{e-ph}$ at the working temperature. The sample holder is connected to the high-frequency port of a bias tee. Dc current is 
applied to the noise source  using a cold current divider connected to the dc port of the bias tee. Finally, the dc-blocked  high frequency port of the bias tee is connected to the port 1 of the JPC via a circulator.

\newpage
\textbf{Supplementary Figures}

      \begin{figure}[h]
\centering
\includegraphics[width=10cm]{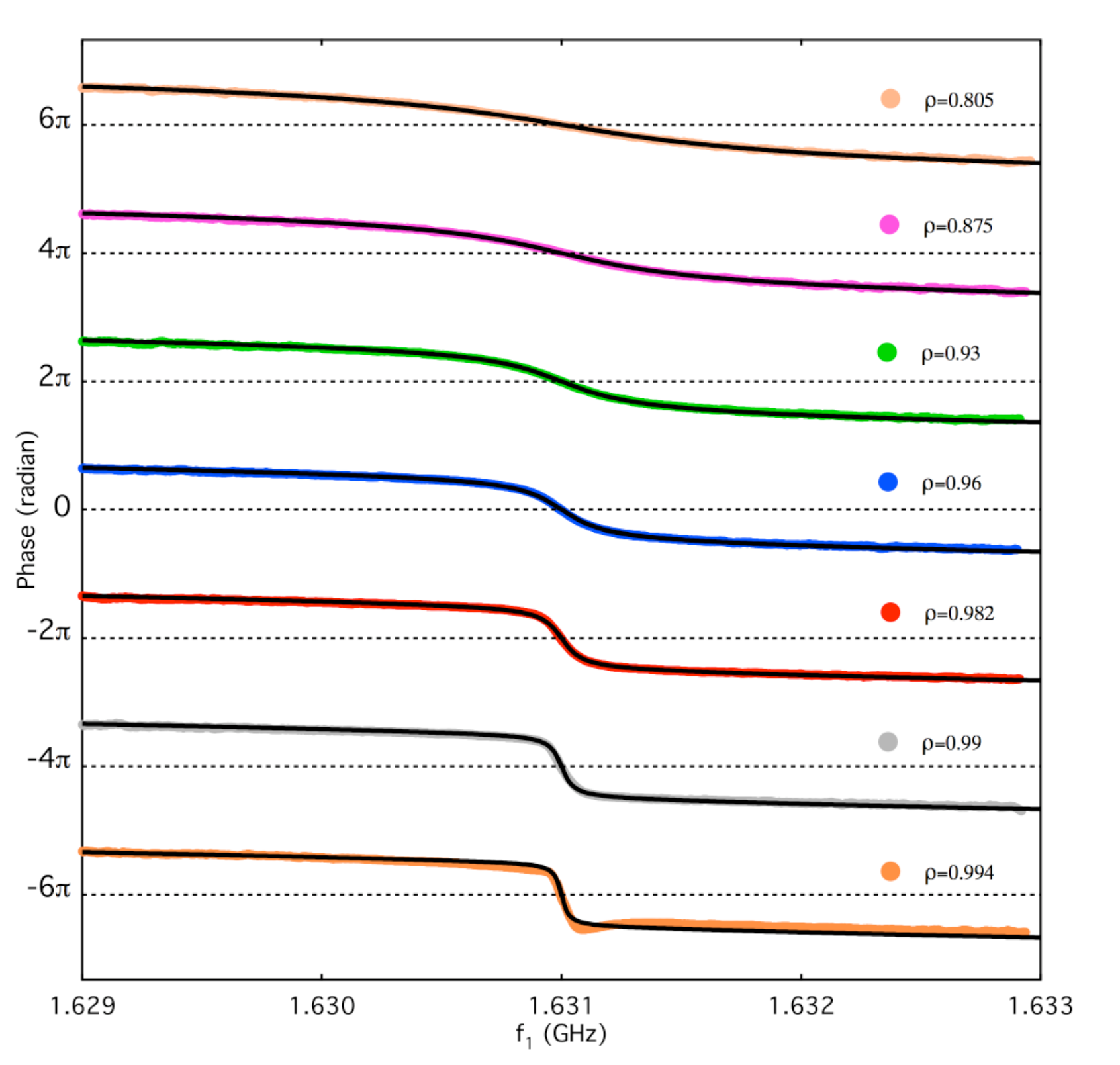}
 \end{figure}
 
 \noindent\small{Supplementary  figure 1. Phase of the ``cis-gain" $r_1$ of the JPC as a function of the drive frequency for different values
 of  the pump power measured at port 1. Solid lines correspond to the theoretical expressions of the phase extracted from expression (\ref{ex_gain}) and obtained for different values of the fit parameter $|\rho|$.}

       \begin{figure}[h!]
\centering
\includegraphics[width=10cm]{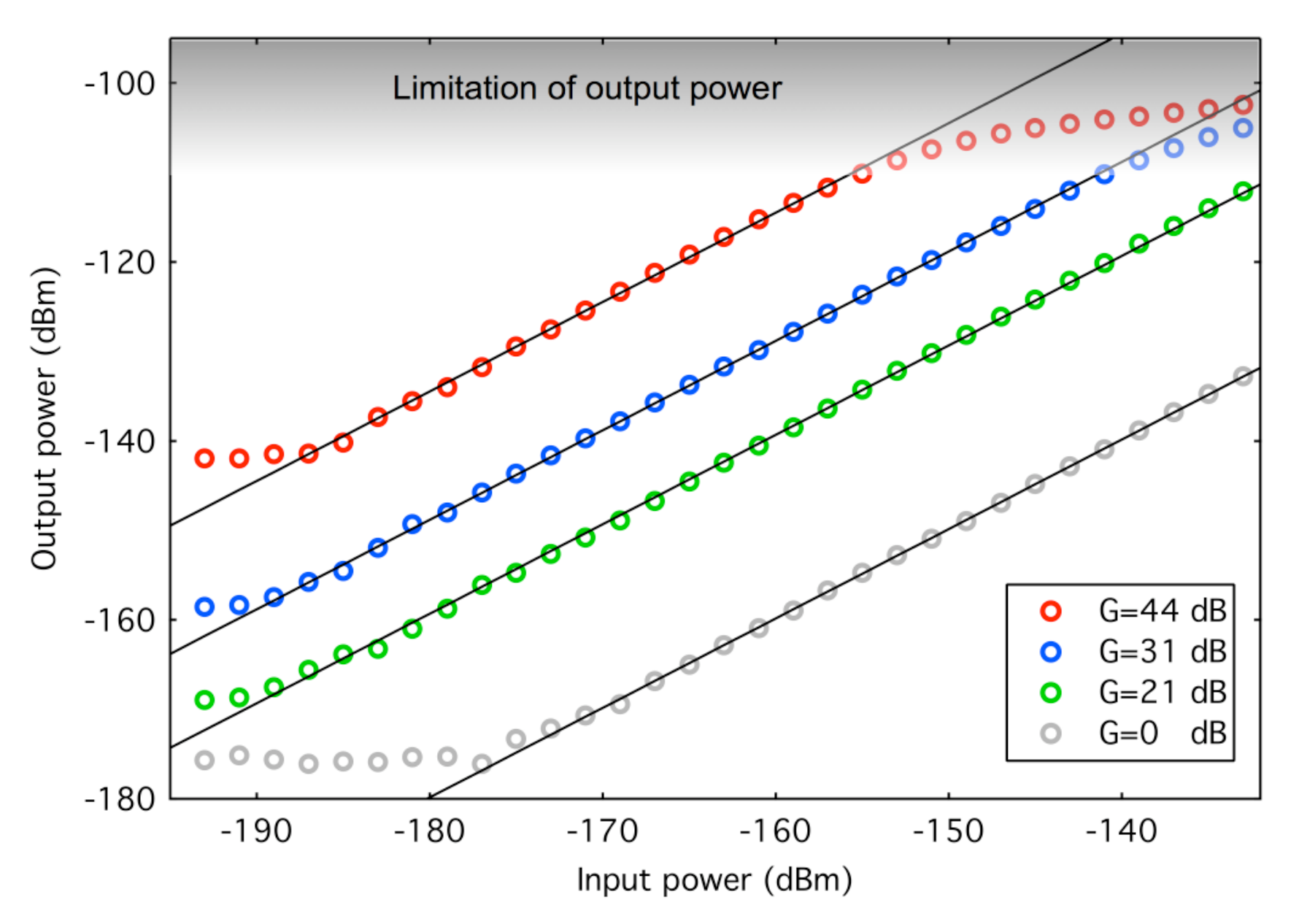}
\end{figure} 

\noindent\small{Supplementary  figure 2. Dynamic range of the amplifier. The figure shows the reflected power at  port 1 of a continuous 
microwave of frequency $f_{1}$=1.631 GHz as a function of its input power for different value of the gain $G$. 
The gain $G$=0 (pump off) is given as reference. The solid lines of slope one show the regime where the gain is independent of the input power.
The curves start to saturate when the output power reach approximatively -105 dB, corresponding to an estimate of the microwave current in the junction taken of order of their critical current  \cite{bergeal}.}

   \begin{figure}[h!]
\centering
\includegraphics[width=8cm]{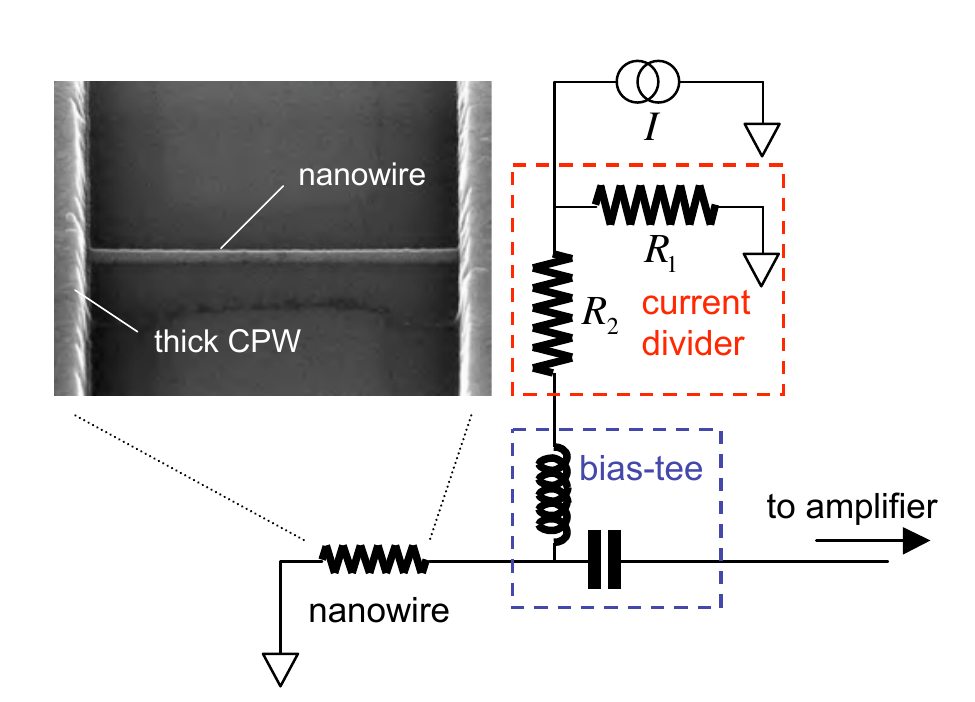}
\end{figure}
\noindent\small{Supplementary figure 3.  Noise measurement set-up and SEM picture of the nanowire embedded in the thick CPW  transmission line. The
  device can be heated by a dc current applied through a current divider at base temperature (resistors $R_{1}$ and $R_{2}$) and a bias tee. The high-frequency port of the bias tee  is connected to the JPC  through a circulator.}

\end{document}